\PassOptionsToPackage{hyphens}{url}
\PassOptionsToPackage{usenames,dvipsnames,svgnames,table}{xcolor}
\documentclass[sigconf,letterpaper,10pt]{acmart}

\usepackage[hyphens]{url}
\usepackage{hyperref}
\usepackage{lipsum}

\usepackage{booktabs} %

\usepackage{etoolbox}
\usepackage{acro}
\usepackage[american]{babel}
\usepackage{pifont}

\hypersetup{breaklinks=true}
\usepackage{epsfig}
\usepackage{caption}
\usepackage{subcaption}
\usepackage{ulem}

\usepackage{colortbl}
\usepackage{units}

\usepackage[inline]{enumitem}
\usepackage{balance}
 \usepackage{array,multirow,graphicx}
 \usepackage{float}
\usepackage{nth}
\usepackage{dblfloatfix}
\usepackage{listings}
\usepackage{tikz}
\usetikzlibrary{automata,backgrounds,fit,positioning,shapes,calc,arrows.meta}

\hypersetup{pdfstartview=FitH,pdfpagelayout=SinglePage}

\makeatletter %

\newcommand{\ie}{i.e.,}
\newcommand{\eg}{e.g.,}
\newcommand{\etal}{et~al\@ifnextchar.{}{.\@}}
\newcommand{\etc}{etc\@ifnextchar.{}{.\@}}

\newcommand{\fig}[1]{Figure~\ref{#1}} 

\newcommand{\afblock}[1]{\noindent{\textbf{#1}}}
\newcommand{\takeaway}[1]{\noindent{\textbf{Takeaway.}} \textit{#1}}

\renewcommand\footnotetextcopyrightpermission[1]{} %

\setcopyright{rightsretained}
\acmConference[]{Technical Report}{RWTH Aachen University}{COMSYS} 

\AtEndDocument{\balance}

\begin{document}
\normalem

\title{Blitz-starting QUIC Connections} 

\renewcommand{\shortauthors}{}

\author{Jan R\"uth}
\email{{rueth, wolsing, serror, wehrle}@comsys.rwth-aachen.de}
\author{Konrad Wolsing}
\author{Martin Serror}
\author{Klaus Wehrle}
\affiliation{%
\institution{RWTH Aachen University}
}

\author{Oliver Hohlfeld}
\email{hohlfeld@comsys.rwth-aachen.de}
\affiliation{%
\institution{Brandenburg University of Technology}
}
\renewcommand{\shortauthors}{R\"uth et al.}

\begin{abstract}
In this paper, we revisit the idea to remove Slow Start from congestion control.
To do so, we build upon the newly gained freedom of transport protocol extendability offered by QUIC to hint bandwidth estimates from a typical web client to a server.
Using this bandwidth estimate, we bootstrap congestion windows of new connections to quickly utilize available bandwidth.
This custom flow initialization removes the common early exit of Slow Start and thus fuels short flow fairness with long-running connections.
Our results indicate that we can drastically reduce flow completion time accepting some losses and thereby an inflated transmission volume.
For example, for a typical DSL client, loading a \unit[2]{MB} YouTube video chunk is accelerated by nearly 2x.
In the worst case, we find an inflation of the transfer volume by 12\% due to losses.

\end{abstract}

\maketitle
\section{Introduction}
\label{sec:intro}
The performance of typical web transfers that are small in size is largely determined by TCP Slow Start~\cite{dukkipati2010}, these flows either never leave Slow Start or spend most of their time in Slow Start.
Slow Start bootstraps congestion control by probing for an appropriate sending rate given that the sender is unaware of the current path conditions to the receiver.
Due to its performance impact, it is subject to discussion since the mid-1980s.
Since then, many improvements and variations to the Slow Start algorithm have been proposed, \eg{} tuning the initial congestion window~\cite{dukkipati2010,floresICDCS2016riptide}, enabling routers to signal available bandwidth~\cite{rfc4782}, or removing it entirely~\cite{jumpstart}.

The deployment of these proposed improvements is challenging in practice.
First, increasing the initial congestion window has been used as means to improve web performance~\cite{dukkipati2010}.
Increasing its size---and thus adapting Slow Start to increasing connection bandwidths---is a discussion ongoing since many decades~\cite{rfc2414,rfc3390,rfc6928,allmanDRAFTabandonIW}.
However, standardized initial window recommendations set an Internet-wide conservative default that is inflexible to address the needs of individual connections.
Second, other attempts require {\em i)} past measurements to signal bandwidth~\cite{floresICDCS2016riptide}, {\em ii)} changes to core routing infrastructures and new signaling mechanisms that hinder deployability~\cite{rfc4782}, {\em iii)} proper signaling means, \eg{} to advertise available bandwidth information~\cite{jumpstart}. 
Apart from Internet-wide defaults for initial window sizes~\cite{ruethIMC17iwv4}, these approaches thus have not been deployed to the Internet given the difficult extendability of TCP.

In this paper, we present \emph{Blitzstart} that revisits the idea of removing Slow Start~\cite{jumpstart} by utilizing QUIC's extendability to hint receiver network conditions to a server for web transactions.
This is relevant since QUIC inherits TCP's congestion control and the associated Slow Start and thus suffers from the same performance penalties as TCP.
Unlike TCP, it now provides means to directly realize powerful signaling mechanisms that we show allow to realize the old concept of removing Slow Start.
This way, we argue that bottlenecks in typical high-performance web applications are either at the sender (\ie{} congestion in a data center) or at the receiver (\ie{} congestion due to slow access links or WiFis) and thus do not require extensive path signaling.
We survey the mechanisms for a receiver to gather bandwidth information that are available today and which would further help congestion control in the future.
Our implementation in Google's QUIC code, evaluated under various network conditions, show the performance gains and fairness with traditional congestion control.
This paper contributes the following:
\vspace{-0.55em}
\begin{itemize}[noitemsep,topsep=5pt,leftmargin=9pt]
	\item A concise analysis of current problems with Slow Start;
	\item Next, we propose how Slow Start can be removed with the help of bandwidth estimates from the client-side;
	\item Our evaluation of four representative connection scenarios shows that with only slight increases in transfer volume, we can significantly reduce the flow completion time; and
	\item Finally, we gain an in-depth understanding on the effects of over- or underestimating the bandwidth in Blitzstart, where a slight overestimation in deeply buffered mobile settings can bring additional benefits.
\end{itemize}
\vspace{-0.5em}
\afblock{Structure.}
Section~\ref{sec:back} revisits Slow Start and congestion control to illustrate its problems.
Section~\ref{sec:design} shows how we can eliminate Slow Start and surveys means for bandwidth estimation (Section~\ref{sec:design:survey}).
We evaluate how our approach performs under various network conditions in Section~\ref{sec:eval}.
Section~\ref{sec:rw} discusses related works and Section~\ref{sec:conclusion} concludes.

\section{Background: Slow Start}
\label{sec:back}
Traditional congestion control implements Slow Start to quickly probe for available bandwidth by effectively doubling the congestion window with each round trip.
It has been shown that this exponential growth is still too slow for many short-lived Internet connections.
In fact, it would be desirable that connections quickly reach a fair bandwidth share, however, in reality, it is known that especially loss-based congestion control such as Cubic often leaves Slow Start prematurely leading to unfairness and bad performance for the new flow~\cite{bbrACMQueue}.
To emphasize this drawback, we investigate the startup behavior of a new short flow competing for bandwidth with a long-lived flow in a simple topology that connects three hosts, two senders attached to the same switch and one receiver attached to another switch, between both switches we have a bottleneck link with a limited capacity of only \unit[50]{Mbit/s}.
We record the network traffic on the receiver side and we plot the rolling average bandwidth over an RTT (configured to \unit[50]{ms}) for each connection.

\begin{figure}
\includegraphics{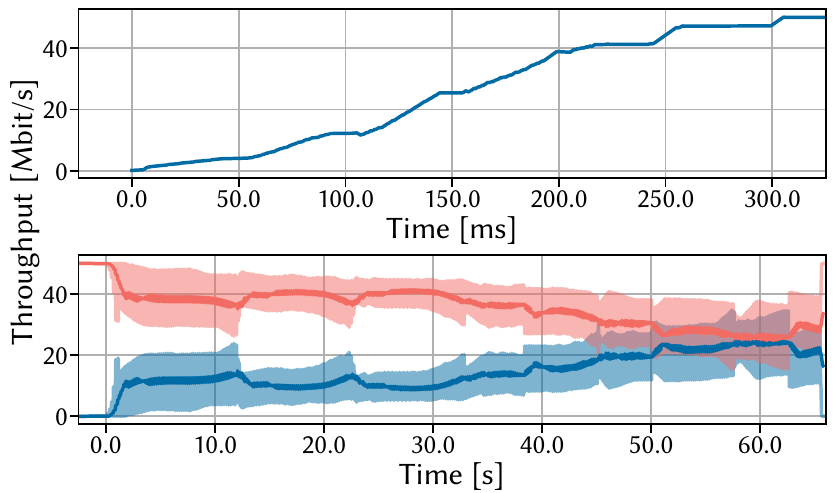}
\caption{Top: Traditional Slow Start of a QUIC Cubic flow. Bottom: Two Cubic flows competing for bandwidth. The blue flow exists Slow Start early, leading to slow fairness convergence and bad performance.}
\label{bg:slowstart}
\end{figure}

The upper plot of \fig{bg:slowstart} shows the startup phase from the first flow when the links are completely idle.
We see the typical exponential growth in Slow Start until roughly \unit[200]{ms}, where Cubic leaves Slow Start and switches to congestion avoidance.
After \unit[300]{ms} the link's capacity is reached.
This leads to reduced performance since, in this example, the whole capacity would be available from the very first RTT.

Looking at the lower plot of \fig{bg:slowstart}, we see a typical Cubic behavior when a second flow (blue) enters while the first flow has already hogged all bandwidth.
Please note that for better visibility, we also plotted (darker color) the average over \unit[1]{s} intervals.
At \unit[0]{s}, the second flow starts with its Slow Start, we can see that it immediately exits Slow Start long before it has reached its fair bandwidth share ($\sim$\unit[25]{Mbit/s}) and, due to congestion avoidance, it takes roughly \unit[60]{s} until a fair bandwidth share is reached.

\takeaway{Slow Start's behavior is especially counterproductive when looking at today's Internet use (\eg{} at homes when using a shared Internet connection), streaming video and downloads fill the bottleneck and comparably short website encounters are never able to reach their fair bandwidth share leading to suboptimal performance.}

Next, we describe how we can reenable an Internet host to quickly gain her fair bandwidth share.

\section{Replacing Slow Start}
\label{sec:design}
The motivation behind Slow Start seems intuitive at first: Congestion control should not induce congestion (and thereby losses) and thus probes the available bandwidth.
Yet, on second thought, all congestion controllers require congestion as a feedback (be it losses or latency increases) to reduce their congestion window and transmission rate.
Thus, by conservatively increasing the transmission rate, we only delay the inevitable: We need to induce congestion to gain bandwidth and thereby also fairness with other flows.
We therefore argue, instead of conservatively increasing the rate and thereby delaying the transmission, to radically push packets into the network (at a reasonable rate) to quickly signal all other congestion controllers the need to reduce their rate.
To this end, we propose to skip Slow Start on connection establishment and directly switch to congestion avoidance, however, we bootstrap the congestion window with a value that seems appropriate given the current network conditions.

To enable a sender to reason about an appropriate value, we signal the used access technology (\eg{} WiFi, DSL, Cable, LTE, ...) as well as a bandwidth estimate (see Section~\ref{sec:design:survey}) to the other end.
With QUIC, this is possible since the packet contents are hidden from the network and thus, this change can be entirely implemented on the end-hosts.
Thus, in a typical web setting, the web browser signals the server a user's access-technology and bandwidth.
From the access technology, a server can reason about buffering, \eg{} WiFi or LTE are known to excessively buffer data to bridge time until airtime is available~\cite{imcLTEbuffers,imcBufferbloatMobile}, and burstiness, \eg{} Cable over DOCSIS connections is known to transmit packets in bursts~\cite{powerboost}, which can help to configure packet pacing at the sender side to meet these burstiness requirements.
With the help of the bandwidth estimate and an RTT sample (which can be derived from a previous connection or an explicit handshake), the congestion window can be estimated as the bandwidth delay product (BDP).
As we will show, the knowledge of buffering can be used to further scale the BDP for optimal connection bootstrap.
Before we dive into the performance gains and downsides of this approach, we survey available technologies for a client to obtain a representative bandwidth estimate.

\subsection{Client-side Bandwidth Estimation}
\label{sec:design:survey}
As we will show, the speed in which we are able to gain a fair bandwidth share depends on a good client-side bandwidth estimation (assuming the bottleneck at the client side).
In this section, we will survey means for a client to establish her bottleneck link speed.
A typical web user can reside in different networks all offering different access technologies.
We are going to discuss three typical scenarios, first, a user at home, connected via Ethernet to the local Internet gateway that connects to the ISP.
Second, the same user using WiFi instead of Ethernet.
Third, a mobile user using her mobile broadband (\eg{} LTE or 3G) to connect to the Internet.

In the first scenario, the local Internet gateway to the ISP is usually the bottleneck.
Even though access speeds have increased over the past years, they are usually inferior to typical gigabit Ethernet that is deployed within a home.
Thus, a client needs to determine the access speed.
For this purpose, she can either leverage past maximum bandwidth observations or rely on existing protocols such as TR-064~\cite{tr064}.
TR-064 offers LAN-side configurability of customer premise equipment (CPE), it can also be used to determine the available link speed to the ISP.
Often additional information is exposed such as the burstiness of the link.

When the user connects via WiFi, depending on the WiFi and access speed, the WiFi can be the bottleneck.
Using the current modulation and coding scheme~(MCS) index, an approximated PHY-rate can be calculated with which an approximate bandwidth can be derived.
Within Linux, this index is queryable from the kernel's wireless subsystem~\cite{nl80211rateinfo}, for OS X, the Core WLAN API~\cite{cwTransmitRate} exists that directly yield the current transmit rate (based on the MCS index) and, on Windows, the Native WiFi API offers similar capabilities~\cite{nativewifiTransmitRate}.

In mobile environments, similar estimates as for WiFi are possible.
For example, for LTE, given the knowledge of the current band, bandwidth and modulation and whether or not MIMO is used, one can calculate the PHY-rate and the overheads that need to be accounted for to derive a transport layer data-rate~\cite{ltetputcalc,3gppltephylayerproc}.
Currently, these calculations are by lack of available information not possible on modern smartphones. 
For Android, the Telephony framework exposes parts of this information but not enough to do the required calculations, for iOS none of this data is publicly accessible.

\begin{table*}[]
\setlength{\tabcolsep}{1.65pt}
\def\arraystretch{0.7}
\setlength\extrarowheight{0.1em}
\begin{tabular}{cl|cc|cc|cc|cc|cc|cc|cc|cc|cc|cc|cc|cc|}
	 &     & \multicolumn{6}{c|}{DSL slow}                                                    & \multicolumn{6}{c|}{DSL fast}                                                    & \multicolumn{6}{c|}{3G}                                                          & \multicolumn{6}{c|}{LTE}                                                         \\[-0.25em]
	 &     & \multicolumn{6}{c|}{\tiny RTT=\unit[50]{ms}, BW=\unit[25]{MBit}, BUF=\unit[50]{ms} (\unit[104]{pkt})}                                                    & \multicolumn{6}{c|}{\tiny RTT=\unit[50]{ms}, BW=\unit[50]{MBit}, BUF=\unit[50]{ms} (\unit[208]{pkt})}                                                    & \multicolumn{6}{c|}{\tiny RTT=\unit[90]{ms}, BW=\unit[8]{MBit}, BUF=\unit[200]{ms} (\unit[140]{pkt})}                                                          & \multicolumn{6}{c|}{\tiny RTT=\unit[70]{ms}, BW=\unit[32]{MBit}, BUF=\unit[200]{ms} (\unit[560]{pkt})}                                                         \\[0.1em]
	 &     & \multicolumn{2}{c|}{70KB} & \multicolumn{2}{c|}{2MB} & \multicolumn{2}{c|}{10MB} & \multicolumn{2}{c|}{70KB} & \multicolumn{2}{c|}{2MB} & \multicolumn{2}{c|}{10MB} & \multicolumn{2}{c|}{70KB} & \multicolumn{2}{c|}{2MB} & \multicolumn{2}{c|}{10MB} & \multicolumn{2}{c|}{70KB} & \multicolumn{2}{c|}{2MB} & \multicolumn{2}{c|}{10MB} \\ \hline
	 &    &    \small FCT         &  \small  Loss           &    \small   FCT       &   \small    Loss      &   \small     FCT      &  \small     Loss       &  \small     FCT       &  \small     Loss        &    \small   FCT       &   \small  Loss         &   \small   FCT        &   \small   Loss        &  \small  FCT           &    \small  Loss         &    \small   FCT       &    \small   Loss      &     \small   FCT      &   \small    Loss       &   \small   FCT        &   \small    Loss       &  \small   FCT        &   \small   Loss       &    \small   FCT       &   \small  Loss    \\[-0.25em]

\multirow{11}{*}{\rotatebox[origin=c]{90}{x BW estimate}}

	 &  &  \tiny x1.3 &  \tiny x4.0 &  \tiny x1.1 &  \tiny x1.5 &  \tiny x1.0 &  \tiny x1.2 &  \tiny x1.0 &  \tiny x1.5 &  \tiny x1.4 &  \tiny x1.3 &  \tiny x1.2 &  \tiny x1.6 &  \tiny x1.1 &  \tiny x3.8 &  \tiny x1.0 &  \tiny x3.5 &  \tiny x1.0 &  \tiny x2.0 &  \tiny x1.8 &  \tiny x1.1 &  \tiny x1.5 &  \tiny x1.9 &  \tiny x1.1 &  \tiny x1.3\\[-0.15em]
	 & 0.5& \textcolor{Orange}{\ding{115}} & \textcolor{Green}{\ding{116}} & \textcolor{Green}{\ding{116}} & \textcolor{Green}{\ding{116}} & \textcolor{gray}{\ding{108}} & \textcolor{Green}{\ding{116}} & \textcolor{gray}{\ding{108}} & \textcolor{Green}{\ding{116}} & \textcolor{Green}{\ding{116}} & \textcolor{gray}{\ding{108}} & \textcolor{Green}{\ding{116}} & \textcolor{BrickRed}{\ding{115}} & \textcolor{gray}{\ding{108}} & \textcolor{Green}{\ding{116}} & \textcolor{gray}{\ding{108}} & \textcolor{Green}{\ding{116}} & \textcolor{gray}{\ding{108}} & \textcolor{Green}{\ding{116}} & \textcolor{Green}{\ding{116}} & \textcolor{gray}{\ding{108}} & \textcolor{Green}{\ding{116}} & \textcolor{gray}{\ding{108}} & \textcolor{gray}{\ding{108}} & \textcolor{gray}{\ding{108}} \\[-0.45em]
	 &  &  \tiny +47ms &  \tiny -13 &  \tiny -170ms &  \tiny -10 &  \tiny -104ms &  \tiny -8 &  \tiny -3ms &  \tiny -3 &  \tiny -407ms &  \tiny +9 &  \tiny -905ms &  \tiny +17 &  \tiny -51ms &  \tiny -16 &  \tiny +454ms &  \tiny -22 &  \tiny +979ms &  \tiny -20 &  \tiny -214ms &  \tiny 0 &  \tiny -1.9s &  \tiny +5 &  \tiny -1.4s &  \tiny -4\\\cline{2-26}
	 &  &  \tiny x1.0 &  \tiny x1.0 &  \tiny x1.4 &  \tiny x3.4 &  \tiny x1.2 &  \tiny x2.7 &  \tiny x1.0 &  \tiny x1.9 &  \tiny x1.9 &  \tiny x7.2 &  \tiny x1.5 &  \tiny x7.4 &  \tiny x1.0 &  \tiny x1.0 &  \tiny x1.3 &  \tiny x1.2 &  \tiny x1.1 &  \tiny x1.2 &  \tiny x1.8 &  \tiny x2.2 &  \tiny x2.5 &  \tiny x9.5 &  \tiny x1.6 &  \tiny x3.1\\[-0.15em]
	 & 1.0& \textcolor{gray}{\ding{108}} & \textcolor{gray}{\ding{108}} & \textcolor{Green}{\ding{116}} & \textcolor{BrickRed}{\ding{115}} & \textcolor{Green}{\ding{116}} & \textcolor{BrickRed}{\ding{115}} & \textcolor{gray}{\ding{108}} & \textcolor{BrickRed}{\ding{115}} & \textcolor{Green}{\ding{116}} & \textcolor{BrickRed}{\ding{115}} & \textcolor{Green}{\ding{116}} & \textcolor{BrickRed}{\ding{115}} & \textcolor{gray}{\ding{108}} & \textcolor{gray}{\ding{108}} & \textcolor{Green}{\ding{116}} & \textcolor{BrickRed}{\ding{115}} & \textcolor{Green}{\ding{116}} & \textcolor{BrickRed}{\ding{115}} & \textcolor{Green}{\ding{116}} & \textcolor{gray}{\ding{108}} & \textcolor{Green}{\ding{116}} & \textcolor{BrickRed}{\ding{115}} & \textcolor{Green}{\ding{116}} & \textcolor{BrickRed}{\ding{115}} \\[-0.45em]
	 &  &  \tiny +2ms &  \tiny 0 &  \tiny -562ms &  \tiny +69 &  \tiny -1.2s &  \tiny +79 &  \tiny +6ms &  \tiny +7 &  \tiny -664ms &  \tiny +176 &  \tiny -1.7s &  \tiny +195 &  \tiny +4ms &  \tiny +1 &  \tiny -2.2s &  \tiny +7 &  \tiny -2.3s &  \tiny +9 &  \tiny -223ms &  \tiny +2 &  \tiny -3.4s &  \tiny +45 &  \tiny -6.4s &  \tiny +33\\\cline{2-26}
	 &  &  \tiny x1.1 &  \tiny x1.3 &  \tiny x1.6 &  \tiny x7.5 &  \tiny x1.3 &  \tiny x5.8 &  \tiny x1.1 &  \tiny x2.2 &  \tiny x2.2 &  \tiny x15.6 &  \tiny x1.7 &  \tiny x15.8 &  \tiny x1.0 &  \tiny x1.3 &  \tiny x1.5 &  \tiny x2.4 &  \tiny x1.2 &  \tiny x2.2 &  \tiny x1.9 &  \tiny x3.2 &  \tiny x3.0 &  \tiny x26.8 &  \tiny x2.1 &  \tiny x8.8\\[-0.15em]
	 & 1.5& \textcolor{Orange}{\ding{115}} & \textcolor{BrickRed}{\ding{115}} & \textcolor{Green}{\ding{116}} & \textcolor{BrickRed}{\ding{115}} & \textcolor{Green}{\ding{116}} & \textcolor{BrickRed}{\ding{115}} & \textcolor{gray}{\ding{108}} & \textcolor{BrickRed}{\ding{115}} & \textcolor{Green}{\ding{116}} & \textcolor{BrickRed}{\ding{115}} & \textcolor{Green}{\ding{116}} & \textcolor{BrickRed}{\ding{115}} & \textcolor{gray}{\ding{108}} & \textcolor{BrickRed}{\ding{115}} & \textcolor{Green}{\ding{116}} & \textcolor{BrickRed}{\ding{115}} & \textcolor{Green}{\ding{116}} & \textcolor{BrickRed}{\ding{115}} & \textcolor{Green}{\ding{116}} & \textcolor{BrickRed}{\ding{115}} & \textcolor{Green}{\ding{116}} & \textcolor{BrickRed}{\ding{115}} & \textcolor{Green}{\ding{116}} & \textcolor{BrickRed}{\ding{115}} \\[-0.45em]
	 &  &  \tiny +19ms &  \tiny +5 &  \tiny -744ms &  \tiny +190 &  \tiny -1.6s &  \tiny +219 &  \tiny +10ms &  \tiny +10 &  \tiny -773ms &  \tiny +412 &  \tiny -2.0s &  \tiny +449 &  \tiny +27ms &  \tiny +7 &  \tiny -3.2s &  \tiny +45 &  \tiny -4.1s &  \tiny +47 &  \tiny -235ms &  \tiny +3 &  \tiny -3.8s &  \tiny +137 &  \tiny -8.7s &  \tiny +122\\\cline{2-26}
	 &  &  \tiny x1.2 &  \tiny x2.0 &  \tiny x1.9 &  \tiny x25.2 &  \tiny x1.4 &  \tiny x17.0 &  \tiny x1.1 &  \tiny x2.9 &  \tiny x2.8 &  \tiny x37.8 &  \tiny x2.1 &  \tiny x42.4 &  \tiny x1.1 &  \tiny x1.8 &  \tiny x1.9 &  \tiny x8.4 &  \tiny x1.3 &  \tiny x6.8 &  \tiny x2.1 &  \tiny x6.5 &  \tiny x3.9 &  \tiny x114.5 &  \tiny x2.9 &  \tiny x44.2\\[-0.15em]
	 & 3.0& \textcolor{Orange}{\ding{115}} & \textcolor{BrickRed}{\ding{115}} & \textcolor{Green}{\ding{116}} & \textcolor{BrickRed}{\ding{115}} & \textcolor{Green}{\ding{116}} & \textcolor{BrickRed}{\ding{115}} & \textcolor{gray}{\ding{108}} & \textcolor{BrickRed}{\ding{115}} & \textcolor{Green}{\ding{116}} & \textcolor{BrickRed}{\ding{115}} & \textcolor{Green}{\ding{116}} & \textcolor{BrickRed}{\ding{115}} & \textcolor{BrickRed}{\ding{115}} & \textcolor{BrickRed}{\ding{115}} & \textcolor{Green}{\ding{116}} & \textcolor{BrickRed}{\ding{115}} & \textcolor{Green}{\ding{116}} & \textcolor{BrickRed}{\ding{115}} & \textcolor{Green}{\ding{116}} & \textcolor{BrickRed}{\ding{115}} & \textcolor{Green}{\ding{116}} & \textcolor{BrickRed}{\ding{115}} & \textcolor{Green}{\ding{116}} & \textcolor{BrickRed}{\ding{115}} \\[-0.45em]
	 &  &  \tiny +40ms &  \tiny +17 &  \tiny -939ms &  \tiny +708 &  \tiny -2.2s &  \tiny +735 &  \tiny +10ms &  \tiny +16 &  \tiny -893ms &  \tiny +1036 &  \tiny -2.7s &  \tiny +1254 &  \tiny +110ms &  \tiny +18 &  \tiny -4.3s &  \tiny +230 &  \tiny -6.7s &  \tiny +233 &  \tiny -259ms &  \tiny +7 &  \tiny -4.2s &  \tiny +602 &  \tiny -10.7s &  \tiny +678\\\cline{2-26}
	 &  &  \tiny x1.3 &  \tiny x2.3 &  \tiny x2.0 &  \tiny x37.2 &  \tiny x1.4 &  \tiny x25.0 &  \tiny x1.1 &  \tiny x3.2 &  \tiny x2.9 &  \tiny x40.5 &  \tiny x2.3 &  \tiny x57.0 &  \tiny x1.1 &  \tiny x1.9 &  \tiny x2.1 &  \tiny x13.1 &  \tiny x1.5 &  \tiny x10.5 &  \tiny x1.8 &  \tiny x7.9 &  \tiny x3.9 &  \tiny x195.1 &  \tiny x3.3 &  \tiny x71.6\\[-0.15em]
	 & 4.0& \textcolor{BrickRed}{\ding{115}} & \textcolor{BrickRed}{\ding{115}} & \textcolor{Green}{\ding{116}} & \textcolor{BrickRed}{\ding{115}} & \textcolor{Green}{\ding{116}} & \textcolor{BrickRed}{\ding{115}} & \textcolor{Orange}{\ding{115}} & \textcolor{BrickRed}{\ding{115}} & \textcolor{Green}{\ding{116}} & \textcolor{BrickRed}{\ding{115}} & \textcolor{Green}{\ding{116}} & \textcolor{BrickRed}{\ding{115}} & \textcolor{BrickRed}{\ding{115}} & \textcolor{BrickRed}{\ding{115}} & \textcolor{Green}{\ding{116}} & \textcolor{BrickRed}{\ding{115}} & \textcolor{Green}{\ding{116}} & \textcolor{BrickRed}{\ding{115}} & \textcolor{Green}{\ding{116}} & \textcolor{BrickRed}{\ding{115}} & \textcolor{Green}{\ding{116}} & \textcolor{BrickRed}{\ding{115}} & \textcolor{Green}{\ding{116}} & \textcolor{BrickRed}{\ding{115}} \\[-0.45em]
	 &  &  \tiny +52ms &  \tiny +23 &  \tiny -998ms &  \tiny +1061 &  \tiny -2.3s &  \tiny +1105 &  \tiny +14ms &  \tiny +18 &  \tiny -915ms &  \tiny +1115 &  \tiny -2.9s &  \tiny +1696 &  \tiny +106ms &  \tiny +19 &  \tiny -4.8s &  \tiny +376 &  \tiny -9.3s &  \tiny +383 &  \tiny -224ms &  \tiny +9 &  \tiny -4.2s &  \tiny +1029 &  \tiny -11.3s &  \tiny +1109\\\hline

\end{tabular}
\caption{Scenarios and mean results over 30 runs compared to standard QUIC. QUIC flow competes for bandwidth. Each row shows our approach under different bandwidth estimates (as a factor to the true BW). Green depicts an improvement (\textcolor{Green}{\ding{115}}), red a deterioration (\textcolor{BrickRed}{\ding{116}}), orange shows FCT changes within the same RTT (\textcolor{Orange}{\ding{115}}\textcolor{Orange}{\ding{116}}) and gray circles depict that the means show no statistically significant difference in the 95\% confidence intervals (\textcolor{Gray}{\ding{108}}).}
\label{tab:overview}
\end{table*}

\section{Evaluation}
\label{sec:eval}
To evaluate the impact of removing Slow Start onto the throughput, packet losses, and fairness in different settings, we utilize Mininet~\cite{mininet}.
While using Mininet offers great flexibility~\cite{ccrMininetReproducibility}, care needs to be taken into its setup to not influence the results.

\afblock{Emulation setup.}
Generally, we want to add delay, buffers, and link speed within our network.
For delay, it is important to not deploy it in the end-hosts, since the (delay induced) buffering within the stack creates a backpressure into the other parts of the stack, altering the behavior of a transport protocol.
We thus implement delays within the core of the network.
Further, buffers for delay need to be sized in accordance to the bandwidth such that adding artificial delay does not result in packet losses.
To model bandwidth, we utilize a token bucket filter with a burst-size of a single packet and set the buffer such that the network is subject to losses and jitter.
We do not add additional jitter, \eg{} from Netem since this leads to packet reordering.

\afblock{Test cases.}
For our tests, we use four hosts, two clients connected to a switch and two servers connected to another switch.
Both switches are connected via a bottleneck link that emulates different link types.
We evaluate our approach under four different settings, a slow DSL link as well as a faster DSL link with a min RTT of \unit[50]{ms}, depicting typical delays to a well-connected web-server.
We further emulate a slow mobile link (3G) that is characterized by a larger min RTT of \unit[90]{ms} and a large buffer of \unit[200]{ms}.
The fourth setting is a faster mobile link (LTE) with a min RTT of \unit[70]{ms} but with the same large buffer.

Table~\ref{tab:overview} summarizes the settings of the four scenarios and gives an overview of our results.
To derive the impact of removing Slow Start and to estimate fairness, we first evaluate the performance of an unmodified QUIC competing for bandwidth with another unmodified QUIC.
This baseline QUIC uses the current 32 segments default as the initial congestion window of which 10 are sent in a burst followed by paced segments over half of the RTT during Slow Start.
Both flows utilize CUBIC for congestion control.
We first run one QUIC flow and let it fully utilize the available bandwidth before we start the QUIC connection that we want to investigate.
Then we measure the flow completion time (FCT) and the packet losses on that QUIC connection and also capture the packets leaving the client's switch to estimate the fairness of both flows.
We transmit \unit[70]{KB}, which was the average size of the compressed Google landing page in 2017, \unit[2]{MB} motivated by the total size of a website according to the HTTPArchive ($>$ \unit[1.5]{MB}) and YouTube video chunks, or \unit[10]{MB} for larger objects such as files or complete videos.

We repeat the same for our modified QUIC version that removes Slow Start and hints a bandwidth estimate to the server.
To show the impact of a good and a bad estimate, we hint 0.5, 1.0, 1.5, 3.0, and 4.0 times the true bottleneck bandwidth to the server.
Since we start our connection in congestion avoidance, we pace packets over 75\% of the RTT just as the regular Google QUIC implementation.

\afblock{Evaluation approach.}
For every setting, we repeat each experiment 30 times and Table~\ref{tab:overview} shows the average increases or decreases in performance in comparison to standard QUIC.
Green arrows denote a performance increase, red arrows a performance decrease, gray circles denote cases where we did not find statistically significant results within the 95\% confidence intervals (we tested for 0 containment in the difference distributions and additionally performed an ANOVA).
Specifically for the FCT, we highlighted results that fall within the same RTT with an orange arrow.

\subsection{Discussion}
\afblock{Correct bandwidth estimate.}
We start our discussion by focussing on the second row of our results (1.0), \ie{} when we correctly estimate the bottleneck bandwidth.
Generally, across all settings, we observe a decrease in FCT except for three cases that show no statistical significance or results that are within the same RTT.
Yet, in cases of the DSL settings, these increases in performance are traded for increased losses.
Please note that these losses are only to be expected once during the lifetime of a connection, and thus pose a one-time overhead.
Still, lost packets must be retransmitted and thus bloat the total transfer size that \eg{} must be transmitted and paid for.
We observe the largest transfer inflation compared to the total transfer size in the DSL fast \unit[2]{MB} case with $\sim$12\%.
Yet, in the same example, we save over 14 round trips in FCT.
In the mobile settings, we again observe increases in performance yet, no significant increases in losses which we attest to the large buffer that we employed.

\afblock{Having a wrong bandwidth estimate.}
Next, we focus on the other rows, \ie{} if we have a wrong bandwidth estimate.
We start by looking at the first row (0.5), in which we assume half the link speed (\ie{} our actual fair bandwidth share).
We again observe mostly performance increases and compared to the previous setting, reduced losses, however as we will show this setting does not yield a fair bandwidth share.

Looking further down the table, we investigate what happens if we overestimate the link speed ($>1.0$)
Generally, we observe further performance increases, however, this time traded for significant increases in lost packets.
Again the DSL fast \unit[2]{MB} setting shows the worst increases in losses, we nearly double the number of bytes that need to be resent.
For the mobile settings, the increases in losses are not that dramatic.
This is due to the large buffering, in fact, it might actually be a good idea to overestimate the bandwidth in these settings.
Since CUBIC congestion control will not operate with an empty buffer but rather fill it over and over again, our estimate derived from the min RTT and the link speed will underestimate the congestion window that CUBIC requires to fill the buffer.

Two more areas require additional attention in the table, namely the \unit[70]{KB} cases for both DSL settings.
Here we observe increased flow completion times and orange arrows.
These \unit[70]{KB} are, in theory, transmittable with two roundtrips (our standard QUIC uses an initial window of 32 segments).
There is a subtle difference for our QUIC and for the regular QUIC operating with Slow Start.
Google QUIC currently paces all traffic but is more aggressive in Slow Start compared to our QUIC which we initialize to congestion avoidance.
In Slow Start, it paces all data over half the RTT and in congestion avoidance over three-quarters of the RTT.
Thus, in the case where we only need two RTTs, our variant will always show slightly reduced FCTs, since it is more cautious in transmitting the data; yet it typically finishes within the same RTT.

\takeaway{
Our results indicate that we are able to trade a slight increase in transfer volume (due retransmitted packet loss) for a significant reduction of flow completion time. 
However, a good bandwidth estimate is the prime requirement.
}

Next, we are going to focus on some of the results that are either characteristic for all our settings or show a significant deviation.

\subsection{Detailed Flow Impact}
In this section, we are going to deep dive into the results of the DSL fast and LTE settings (slow and 3G are similar, respectively).
\begin{figure}
\includegraphics{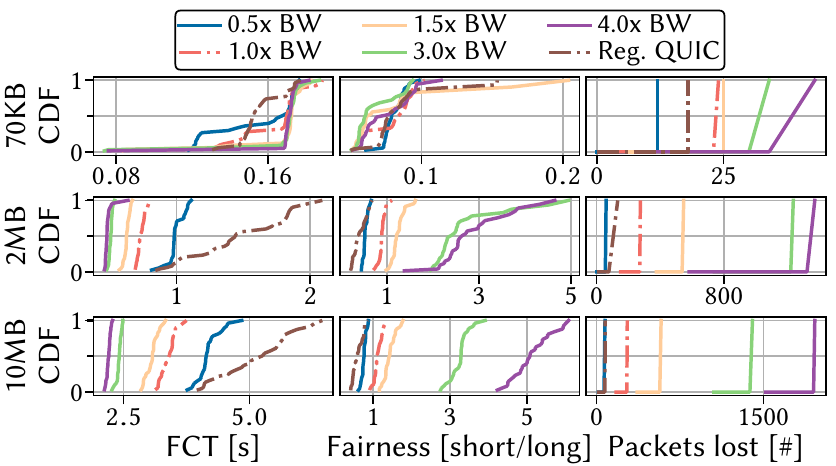}
\caption{Detailed performance of the DSL fast scenario. Plots show CDFs of, left: FCT, middle: fairness as ratio of bytes transmitted while short flow was active, and, right: packet losses. Each row depicts results for different transfer sizes.}
\label{fig:detailDSLfast}
\end{figure}
\begin{figure}
\includegraphics{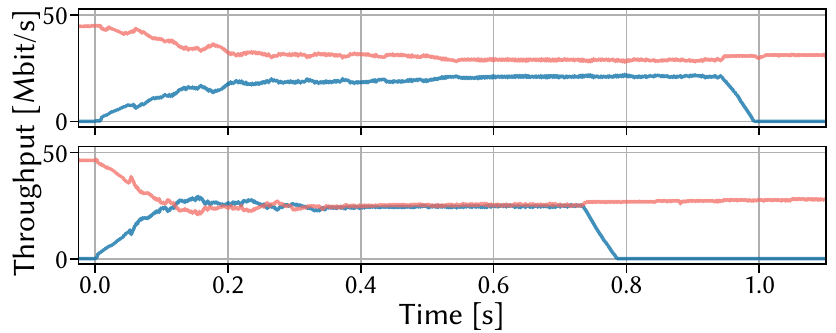}
\caption{Example connection start from  \unit[2]{MB} DSL fast setting. Top, default behavior with Slow Start when things work out (fast part of the FCT distribution). Bottom with Blitzstart, example taken from the middle of the FCT distribution. Blitzstart finishes faster and is more fair with the existing flow.}
\label{fig:medianstartup}
\end{figure}

\afblock{DSL setting.}
\fig{fig:detailDSLfast} shows the former setting in greater detail.
Each row in the figure represents one transmission size.
The columns show the FCT, the fairness and the lost packets as CDFs over 30 individual measurements runs.
To compute the fairness, we divide the sum of bytes of the short and the long flow, starting from when the short flow enters until it finishes.
If this ratio is below one, the long flow transmitted more bytes and when above one, the fairness is skewed towards the short flow that transmitted more bytes.

The \unit[70]{KB} row highlights our previous observations from Table~\ref{tab:overview} since there is no significant difference between all tested settings.
Especially, we find only marginal differences in the FCT tail of the distribution.
Further, the number of bytes and RTTs over which these are transmitted are simply too small to steal bandwidth from the long flow leading to a fairness imbalance towards the long flow.

\begin{figure}
\includegraphics{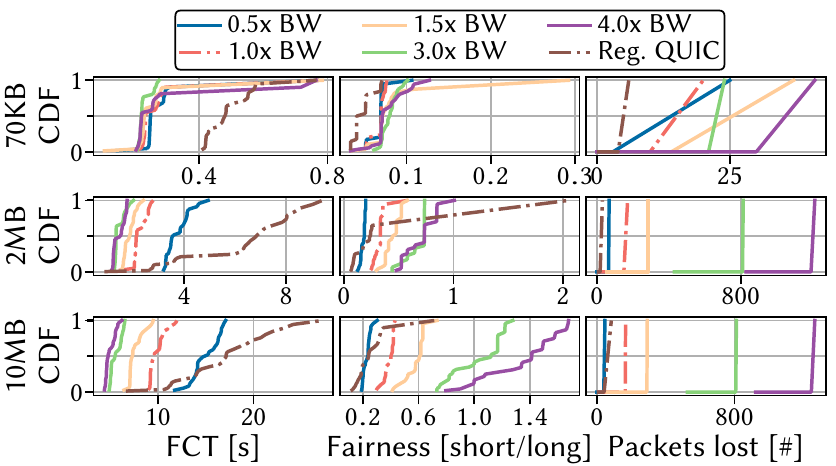}
\caption{Detailed performance of the deep buffered LTE scenario. Plots show CDFs of, left: FCT, middle: fairness as ratio of bytes transmitted while short flow was active, and, right: packet losses. Each row depicts results for different transfer sizes.}
\label{fig:detailLTE}
\end{figure}

Looking one row further down (\unit[2]{MB}) and comparing the 1.0x and regular QUIC CDFs (dash-dotted lines), we observe that the FCT distributions are significantly different.
All bandwidth estimated variants show a steep incline in contrast to the regular QUIC variant that suffers from varying performance and drastic worsening at the tail of the distribution.
This is also reflected in the distribution of fairness ratios, where regular QUIC is not able to quickly gain fairness with the long-running flow.
\fig{fig:medianstartup} reassures this impression (please note that the visually sloped start and sloped termination are effects of the rolling mean), on the top we see one of the fastest regular QUIC flows, the blue short flow requires a long time to reach an equilibrium.
Even then, there is a significant performance advantage for the long flow.
In contrast to this, the lower plot shows one Blitzstart run from the median of the distribution, after $\sim$\unit[100]{ms} (2 RTTs) we reach a comparable throughput which both flows keep for the lifetime of the short flow.
Interestingly, this behavior is all reached by a very constant overhead of lost packets that shows only little variation and only slight increases when the correct bandwidth is estimated.

The \unit[10]{MB} shows a very similar picture to the \unit[2]{MB} case.
For all settings, we see that the drastic performance increases of the 3.0x and 4.0x cases are only achieved by drastic losses resulting in extremely unfair congestion control.
We repeated our measurements for a smaller buffer size (\unit[5]{ms}) and we could show the same effects with slightly reduced overall throughput in all settings.

\afblock{LTE setting.}
\fig{fig:detailLTE} shows the same evaluation for the deeply buffered setting.
While again fairness cannot be achieved in the \unit[70]{KB} setting, all Blitzstarted variants show performance increases in the majority of all measurements.
This is no surprise, the large buffer induces a large delay for the short flow until it is able to increase the congestion window in Slow Start.
The blitzstart variants push more packets into the network at the start thus, avoiding overly long stalls, at the cost of slightly increased losses.
However, we observe that there are still cases at the tail of the distribution in which both variants perform similarly.

When increasing the transfer size to \unit[2]{MB}, the 0.5x estimated variant loses significant ground to the others.
Compared to the DSL setting, we have troubles reaching a fair bandwidth share, we believe this is due to the large buffer which delays congestion feedback and the transfer is completed before enough feedback is delivered to the long flow.
In fact, it might actually be beneficial to overestimate the bandwidth.
Since our Blitzstart bases on an observed min RTT and the BW estimate to calculate the congestion window at startup, for a Cubic congestion control, it underestimates the maximum congestion window that would be given with a filled buffer.
Our experimentations even showed that even if we keep the newly starting flow running, it will never reach a fair equilibrium with the already started one.

Looking at the \unit[10]{MB} setting, we can see that by 3.0x overestimating the BW, we are to come close to a fair BW share.
Yet, the fairness CDF also shows that this is not reliably reproducible.
Further, we need to induce heavy losses to achieve this fairness resulting in an immense inflation of the transfer size.
Yet, overestimations like 1.5x still show acceptable losses and further increased performance.
Since Blitzstart is informed about access technology, it enables a congestion control to reason about sensible overestimations.
Further, not yet explored, directions could fine-tune more aggressive pacing rates in the first couple of round trips to gain a fairer bandwidth share with an established flow.

\takeaway{
Our results again highlight the current state of affairs, short flows are significantly disadvantaged.
Blitzstart helps to reestablish fairness, we find that in heavily buffered settings, a slight overestimation of the bandwidth estimate can further help to quickly bootstrap the short flow.}

\section{Related Work}
\label{sec:rw}

Removing Slow Start from congestion control is not a new idea.
To this end, Quickstart~\cite{rfc4782} suggests to remove Slow Start and enable routers on the path to signal available bandwidth.
Since this requires network changes, no deployment in the Internet is known.
Jumpstart~\cite{jumpstart} requires sender-only changes and is the foundation for this work.
There, a sender simply paces all available packets during connection startup being only limited by the receive window.
In contrast to TCP, QUIC paces by default but enables Internet-wide signaling without having to worry principle deployability.
Further, our study evaluates the Jumpstart idea in Google QUIC thereby shedding light on its performance beyond simulations in TCP.
Halfback~\cite{halfback} extends Jumpstart by proactively retransmitting the tail of the congestion window thereby boosting flow completion times when packets are lost.
However, this comes at the price of an unknown transfer size inflation which we deem infeasible, we rather rely on QUIC's RACK-style~\cite{tcpRACK} fast recovery.
Setting larger initial congestion windows~\cite{dukkipati2010} has seen Internet-wide deployment~\cite{ruethIMC17iwv4}.
CDNs are also known to tune their initial congestion windows, \cite{ruethTMAiwcdn}~finds CDNs to use initial windows in the range with which Google QUIC experiments (by default 32 segments, 50 segments configurable).
This rather inflexible static assignment is tackled by Riptide~\cite{floresICDCS2016riptide} and SmartIW~\cite{smartIW}, which learn initial windows from past connections.

While past connections can hint at receiver network conditions, we rather see them as a building block that can be used in combination with a receiver-side bandwidth hint.
In our work, we highlighted possible angles in which to tackle bandwidth estimates but do not go into detail.
\cite{Lu2015cqic} shows how to utilize cellular (4G) PHY-layer information at the client for bandwidth estimations, they even implement it in the Google QUIC framework but do not show the effects on other flows, \eg{} fairness and losses nor focus on the startup phase.
Further bandwidth estimations~\cite{samba2016throughputcellular,bui2014throughputmobile} utilize client context information (\eg{} location, signal quality, speed) in cellular networks to learn and predict channel throughput, which could be used in an implementation of our system.

\section{Conclusion}
\label{sec:conclusion}
In this paper, we address the well-known problems of Slow Start in congestion control by bootstrapping congestion windows of typically short-lived web transactions using bandwidth estimates from the client.
We revisit the idea of removing Slow Start which has suffered from deployable bandwidth estimates so far.
Fueled by QUIC's extendability, our implementation in Google QUIC hints receiver access technology and network conditions to a server and thus offers a viable and \emph{deployable} solution.
In our evaluation, we consider four representative scenarios, showing that a small increase in transfer volume, due to slightly higher packet losses, is traded against a significant reduction of the flow completion time.
We thereby help short flows to regain fairness with bandwidth-hogging long flows.
Moreover, an in-depth analysis when under- or overestimating the bandwidth emphasize the importance of accurate estimations, while also revealing that a slight overestimation shows an increased performance in deeply buffered settings.
While our work shows promising results, future work should evaluate Blitzstart's impact when more connections are present, other connection race Blitzstart at the very start of the connection as well as Blitzstart's impact, interplay, or utilization within other congestion controllers such as BBR.

\section{Paper Origin \& Meta Discussion}
We originally submitted this paper to the CoNEXT EPIQ 2018\footnote{\url{https://conferences2.sigcomm.org/co-next/2018/\#!/workshop-epiq}} workshop where it was not selected for publication.
Subsequently, we presented and discussed results of this paper at the QUIC Symposium 2019 in Diepenbeek, Belgium\footnote{\url{https://quic.edm.uhasselt.be/symposium19/}}.
The attendants largely expressed interest in the presented paper, which motivates us to share it with the community with an extended discussion of the reviewers feedback.

In the following, we provide a summary of the reviewer's comments (with a focus on the technical aspects) on the paper to allow other researchers and the community to build upon our work.
We will also briefly comment on these reviewer's input, \eg{} in what way we believe those should be tackled and what their challenge is.

\afblock{Concern: The paper investigates only two flows.}\\
\textit{Summary:} In the paper, we compare only a single elephant flow against a newly arriving flow.
It is unclear how the approach behaves when, \eg{} multiple short flows with the same bandwidth estimate arrive at the same time (\eg{} when a bottleneck is shared by multiple users) or shortly after each other, \eg{} when resources (requiring new connection) are discovered on a website. 
It is further unclear how non-blitzstarted short flows behave when entering a bottleneck simultaneously.
Moreover, it is unclear what happens if multiple long flows already competing for bandwidth and a new blitzstarted flow behave.
It could also be interesting to investigate the losses for the already running flow.\\
\textit{Comments:} While the scope of the paper was to motivate the general idea of achieving performance improvements by announcing bandwidth estimates, the detailed evaluation is beyond the scope of this concept paper. There are a number of challenges involved that can be picked-up by follow up work. A challenge when emulating multiple flows is that those flows must be independent of each other and not synchronize otherwise the insights into how congestion control behaves can be misleading.
In the case of multiple flows from the same host are blitzstarted, what bandwidth estimates should it announce? A fair fraction?
In the same case it could be interesting to actively signal ongoing flows to reduce their rate to reduce losses and to signal newly available bandwidth once a flow departs.

\afblock{Concern: Bandwidth estimates could be a fingerprinting vector.}\\
\textit{Summary:} The bandwidth estimates that we require for our approach could be used for fingerprinting or tracking a user.
Since the availability of bandwidth could be unique to certain users or show certain patterns, \eg{} when being mobile, it could allow a powerful observer to track users.
To counter such tracking, one could think of quantizing the bandwidth information, it could be interesting to see how such a quantization affects the performance and how coarse such a quantization must be.\\
\textit{Comment:} Our primary motivation addressed by this paper is to show that providing bandwidth estimates can improve performance.
The question of how available bandwidth is unique and allows fingerprinting goes way beyond the paper's scope of showing its principle performance benefits.
Nevertheless, the proposed quantization offers an interesting angle for anonymization that can be investigated in terms of how performance and coarseness can be traded.

\afblock{Concern: Bandwidth estimates are not convincing.}\\
\textit{Summary:} The paper only briefly surveys methods to estimate the bandwidth but does not quantify their accuracy or availability.
Given the large number of different CPE devices, how well are such measurements supported and how accurate are they?
In general, the reviewer's were skeptical if determining the bandwidth and access technology is reliably possible.\\
\textit{Comment:} Since we did not investigate the accuracy or availability, we cannot comment on this.
We, however remark that past measurements could still be used to bootstrap the connection, those could even be only allowed for subsequent connections and be authenticated similar to 0-RTT connection attempts.

\afblock{Concern: Congestion collapse and false bandwidth estimates.}\\
\textit{Summary:} False estimates and multiple flows (see first concern) could lead to a congestion collapse.
Participants cheating the system, \eg{} by announcing too large rates, could perform denial of service attacks towards certain links.\\
\textit{Comment:} This is a valid concern and should be in part addressed by investigating how multiple flows using this technique behave.
Further mechanisms to validate reasonable bandwidth announcements could be made on the server side, \eg{} comparing requests to past maximum delivery rates or capping the total bandwidth towards certain hosts.

\afblock{Concern: Bandwidth estimates could be exchanged outside of the transport layer.}\\
\textit{Summary:} Other cross-layer designs could be investigated, it is not a requirement to transport this information within QUIC.\\
\textit{Comment:} While it is certainly possible to do this outside the transport, \eg{} on a HTTP request. 
This would require significant instrumentation of the transport and cross-layer interfaces. 
We feel it rightfully belongs to the transport.

Apart from these concerns, some reviewer's remarked that they would have preferred a greater focus on the shortcomings of the approach.

\bibliographystyle{ACM-Reference-Format}
\balance
\bibliography{literature} 

\end{document}